\documentstyle[twoside,fleqn,npb,epsfig]{article}
\newcommand{\AmS}{{\protect\the\textfont2
  A\kern-.1667em\lower.5ex\hbox{M}\kern-.125emS}}

\title{Inclusive Jet Cross Sections in $\overline{p}p$ Collisions at 
$\sqrt{s}$= 630 and 1800~GeV}

\author{V.\ Daniel Elvira\address{Department of Physics and Astronomy,
        State University of New York at Stony Brook, Stony Brook,\\
        NY 11794-3800}
        \thanks{Representing the D\O\ Collaboration}}
       
\begin{document}

\begin{abstract}
We have made a precise
measurement of the inclusive jet cross section at $\sqrt{s}$=1800~GeV.
The result is based on an integrated luminosity of 92 pb$^{-1}$ 
collected at the Fermilab Tevatron $\overline{p}p$ Collider with the
D\O\ detector. The measurement is reported as a function of
jet transverse energy (60~GeV $\leq E_{T} \leq$ 500~GeV ), and in the 
pseudorapidity intervals $|\eta|\leq$ 0.5 and 0.1$\leq |\eta| \leq$0.7.
A preliminary measurement of the pseudorapidity dependence of inclusive jet 
production ($|\eta|\leq$ 1.5 ) is also discussed.
The results are in good agreement with predictions from 
next--to--leading order (NLO) quantum chromodynamics (QCD). 
D\O\ has also determined the ratio of jet 
cross sections at $\sqrt{s}$=630~GeV and $\sqrt{s}$=1800~GeV 
($|\eta|\leq 0.5$). This preliminary measurement differs from 
NLO QCD predictions.
\end{abstract}

\maketitle

\section{Introduction}

Within the framework of quantum chromodynamics (QCD), inelastic scattering 
between a proton and antiproton is described as a hard collision between 
their constituents (partons). After the 
collision, the outgoing partons manifest themselves as 
localized streams of particles or ``jets''. 
Predictions for the inclusive jet cross section have 
improved in the early nineties with next-to-leading order (NLO) perturbative
QCD calculations
\cite{theory} and new, accurately
measured parton density functions (pdf)\cite{pdfs}.

D\O\ has recently measured and published~\cite{d0paper} the cross section for 
the production of jets as a function of the jet energy transverse to the 
incident beams, $E_{T}$. The measurement
is based on an integrated 
luminosity of about 92~pb$^{-1}$ of $\overline{p}p$ hard 
collisions collected with
the D\O\ Detector~\cite{D0detector} at the Fermilab Tevatron 
Collider.
This result allows a stringent test of QCD, with a total uncertainty
substantially reduced relative to previous results~\cite{old,nCDF}.  
We also measure the ratio of jet cross sections at two center-of-mass
energies: 630 (based on an integrated luminosity of about 0.537 pb$^{-1}$)
and 1800~GeV~\cite{dpf}. Experimental and theoretical uncertainties
are significantly reduced in the ratio. This is due to the large 
correlation in the errors of the two cross section measurements,
and the suppression of the sensitivity to parton distribution functions (pdf) 
in the prediction. The ratio of cross sections thus provides a stronger
test of the matrix element portion of the calculation than a single cross
section measurement alone. 
Previous measurements of cross section ratios 
have been performed with smaller data sets by the UA2 and 
CDF~\cite{ratio} experiments.

\section{Jet Reconstruction and Data Selection}

Jets are reconstructed using an iterative jet cone algorithm
with a cone radius of $\cal{R}$=0.7 in $\eta$--$\phi$ space, 
(pseudorapidity is defined as 
$\eta = -{\rm ln}[{\rm tan}\frac{\theta}{2}]$)~\cite{d0the}.  
The offline data selection procedure, which eliminates background caused by
electrons, photons, noise, or cosmic rays, follows the methods described in
Refs.~\cite{dan,krane}. 

\section{Energy Corrections}

The jet energy scale correction, described in \cite{escale}, removes
instrumentation effects associated with calorimeter response, showering, and
noise, as well as the contribution from spectator partons (underlying event).

The energy scale corrects jets from their reconstructed 
$E_{T}$ to their ``true'' $E_{T}$ on average. An unsmearing correction is
applied later to remove the effect of a finite $E_{T}$ 
resolution~\cite{d0paper}.  

\section{The Inclusive Jet Cross Section}

The resulting inclusive double differential 
jet cross sections, $\langle d^2\sigma / (dE_{T} d\eta) 
\rangle$, for $|\eta| \leq 0.5$ and $0.1 \leq |\eta| \leq 0.7$ (the
second region for comparison to Ref.~\cite{nCDF}), 
are compared 
with a NLO QCD theoretical prediction~\cite{theory}.
Discussions on the different choices in the theoretical calculation:
pdfs, renormalization and factorization scales ($\mu$), and 
clustering algorithm
parameter ($R_{sep}$) can be found in Refs.~\cite{d0the}.  

\begin{figure}[htb]
\begin{minipage}[t]{3.05in}
{\centerline{\psfig{figure=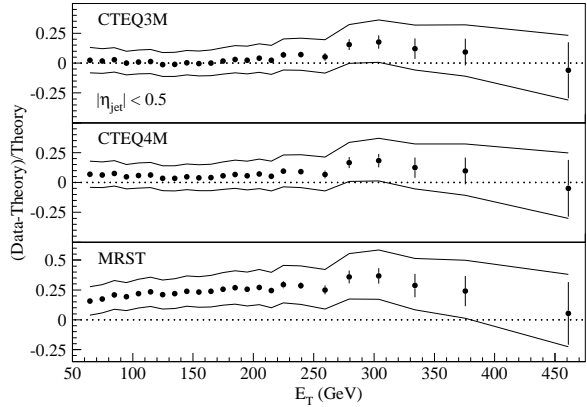,width=3.05in}}}
\vskip-1cm
\caption{The difference between data and {\small JETRAD} QCD predictions
    normalized to predictions.  The bands are the total experimental 
uncertainty.}
\vskip-0.8cm
  \label{Fig_3}
\end{minipage}
\hspace*{2mm}
\end{figure}

Figure \ref{Fig_3} shows the ratios $(D-T)/T$ for the data ($D$) and
{\small JETRAD} NLO theoretical ($T$) predictions based on the CTEQ3M, CTEQ4M
and MRST pdf's [4,5] for $|\eta| \leq 0.5$.  
(The tabulated data for both $|\eta| \leq 0.5$ and $0.1 \leq |\eta| \leq 0.7$
measurements can be found in Ref.~\cite{matrix}.)

The predictions are in good quantitative agreement with the data, as
verified with a
$\chi^{2} = \sum_{i,j} (D_{i}-T_{i}) (C^{-1})_{ij} (D_{j}-T_{j})$
test, which incorporates the uncertainty covariance matrix
$C$. Here $D_{i}$ and $T_{i}$ represent the $i$-th data
and theory points, respectively. The overall systematic uncertainty is 
largely correlated.

Table~\ref{tab:table2} lists $\chi^{2}$ values for several {\small JETRAD}
predictions using various parton distribution 
functions~\cite{pdfs}.
The predictions describe both the $|\eta| \leq 0.5$ and 
$0.1 \leq |\eta| \leq 0.7$ cross section 
very well.
The measurement by D\O\ and CDF are also in good quantitative 
agreement within their systematic uncertainties~\cite{d0paper}.

\begin{table}[htbp]
\vskip-0.8cm
\caption{ $\chi^{2}$ comparisons between {\small JETRAD} and $|\eta| \leq 0.5 $
and $0.1 \leq |\eta| \leq 0.7 $ data for $\mu = 0.5E_{T}^{\rm max}$,
$\cal{R}_{\rm{sep}}$=$1.3\cal{R}$, and various
pdfs.  There are 24 degrees of freedom. }
\begin{tabular}{c r r}
pdf &  $|\eta| \leq 0.5 $ &  $0.1 \leq |\eta| \leq 0.7 $ \\ \hline
  CTEQ3M     &  23.9     &  28.4    \\
  CTEQ4M     &  17.6     &  23.3    \\
  CTEQ4HJ    &  15.7     &  20.5    \\
  MRSA\'     &  20.0     &  27.8    \\
  MRST       &  17.0     &  19.5    \\
\end{tabular}
\vskip-1.2cm
\label{tab:table2}
\end{table}

\vskip-2cm

\section{$\eta$ Dependence of 
$\langle d^2\sigma / (dE_{T} d\eta) \rangle$}

\begin{figure}[htb]
%\vskip-1.2cm
\begin{minipage}[t]{3.05in}
{\centerline{\psfig{figure=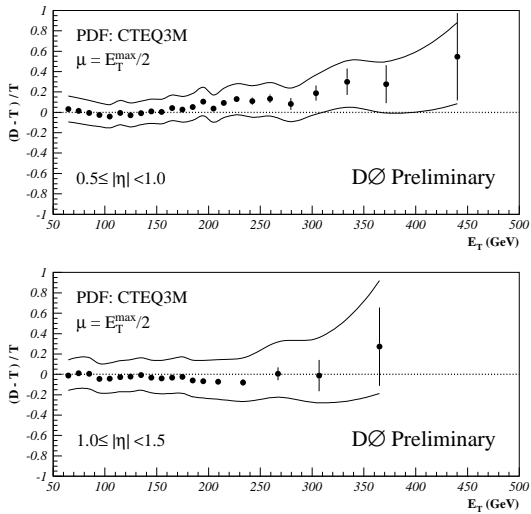,width=3.05in}}}
\vskip-1.1cm  
\caption{Pseudorapidity dependence of the inclusive jet cross section 
(0.5$<|\eta|<$1 and 1$<|\eta|<$1.5). Comparison between data and NLO QCD
predictions. The bands are the total systematic uncertainty in the
experiment.}
  \label{fig:forward}
\end{minipage}
\vskip-0.5cm
\hspace*{2mm}
\end{figure}

D\O\ has made a preliminary measurement of the pseudorapidity 
dependence of the inclusive jet cross
section. Figure~\ref{fig:forward} shows 
the ratios $(D-T)/T$ for the data ($D$) and
{\small JETRAD} NLO theoretical ($T$) predictions using the CTEQ3M pdf set
for $0.5 \leq |\eta| \leq 1$ and $1 \leq |\eta| \leq 1.5$.   
The measurements and the predictions are in good qualitative agreement.
A detailed error analysis is currently being completed.

\section{Ratio of Scale Invariant Jet Cross Sections}

\begin{figure}[htb]
\begin{minipage}[t]{3.05in}
{\centerline{\psfig{figure=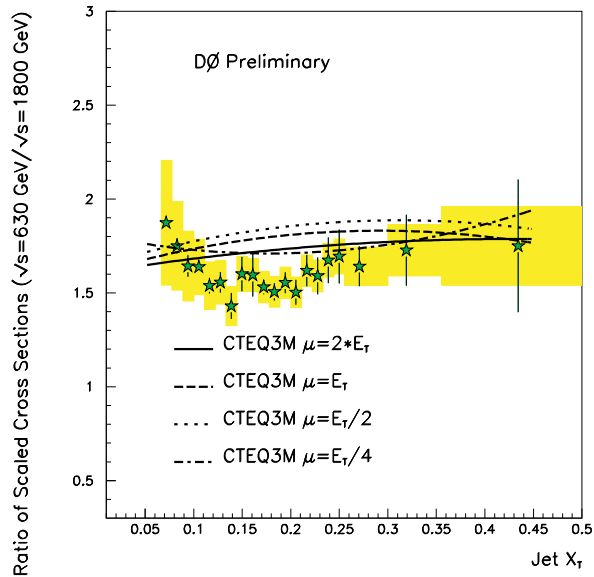,width=3.05in}}}
\vskip-1cm
\caption{The ratio of scale invariant jet cross sections.
The stars are the D\O\ data, the band is the systematic uncertainty, 
and the lines are the
NLO QCD predictions.}
\vskip-0.4cm
\label{fig:ratmu}
\end{minipage}
\hspace*{2mm}
\end{figure}

A simple parton model would predict a jet cross section that scales with
center-of-mass energy. In this scenario, $E_{T}^{4} \cdot E 
\frac{d^{3}\sigma}{dp^{3}}$, plotted as a function of jet $x_{T}\equiv
\frac{2 \, E_{T}}{\sqrt{s}}$, would remain constant with respect to the
center-of-mass energy. Figure~\ref{fig:ratmu} shows the D\O\ measurement
of $E_{T}^{4} \cdot E \frac{d^{3}\sigma}{dp^{3}}$ (stars) compared to
{\small JETRAD} predictions (lines). There is poor agreement between data
and NLO QCD calculations using the same $\mu$ in the numerator and
the denominator (probability of agreement not greater than 10\%). 
The agreement improves for predictions with different $\mu$ at the two
center-of-mass energies~\cite{dpf}.

\vskip0.3cm

In conclusion, we have made precise measurements of jet production
cross sections. At $\sqrt{s}$=1800~GeV, there is good agreement between
the measurements and the data.
The ratio of cross sections at $\sqrt{s}$=1800 and 630~GeV, however,
differs from NLO QCD predictions, unless different renormalization scales
are introduced for the two center-of-mass energies.

\end{document}